# A Low Power Threshold, Ultrathin Optical Limiter Based on a Nonlinear Zone Plate

YUQI ZHAO,[1,2] HAMIDREZA CHALABI,[1,2] AND EDO WAKS[1,2,3,*]

[1]Department of Electrical and Computer Engineering, University of Maryland, College Park, Maryland, USA, 20742
[2]Institute for Research in Electronic and Applied Physics (IREAP), University of Maryland, College Park, Maryland, USA, 20742
[3]Joint Quantum Institute (JQI), University of Maryland, College Park, Maryland, USA, 20742
*edowaks@umd.edu

**Abstract**: Ultrathin optical limiters are needed to protect light sensitive components in miniaturized optical systems. However, it has proven challenging to achieve a sufficiently low optical limiting threshold. In this work, we theoretically show that an ultrathin optical limiter with low threshold intensity can be realized using a nonlinear zone plate. The zone plate is embedded with nonlinear saturable absorbing materials that allow the device to focus low intensity light, while high intensity light is transmitted as a plane wave without a focal spot. Based on this proposed mechanism, we use the finite-difference time-domain method to computationally design a zone plate embedded with InAs quantum dots as the saturable absorbing material. The device has a thickness of just 0.5 $\mu$m and exhibits good optical limiting behavior with a threshold intensity as low as 0.45 kW/cm$^2$, which is several orders of magnitude lower than current ultrathin flat-optics-based optical limiters. This design can be optimized for different operating wavelengths and threshold intensities by using different saturable absorbing materials. Additionally, the diameter and focal length of the nonlinear zone plate can be easily adjusted to fit different systems and applications. Due to its flexible design, low power threshold, and ultrathin thickness, this optical limiting concept may be promising for application in miniaturized optical systems.

## 1. Introduction

Optical limiters protect light sensitive components by transmitting low intensity light while attenuating high irradiance above a certain power threshold. With the development of miniaturized optical systems, there is a need for ultrathin optical limiters. However, classical optical limiters are generally millimeter-thick devices made of bulk materials or suspensions with high limiting threshold intensities and have therefore been insufficient for the needs of miniaturized optical systems [1–3].

Flat optics, such as metasurfaces, provide a promising approach for achieving ultrathin optical limiters [4–11]. However, most of current metasurfaces-based optical limiters rely on a relatively weak third-order nonlinear effect, which results in high power limiting thresholds of $10^2$–$10^7$ kW/cm$^2$ [4-9]. Recently, a new method incorporating metasurfaces with phase-change materials was proposed to realize optical limiting with a lower threshold at 3kW/cm$^2$ [10,11]. To further decrease the required threshold intensity of an ultrathin optical limiter, we need to explore flat optics that incorporate different nonlinear mechanisms. For example, zone plates are another type of flat optic, which consist of radially symmetric zones that alternate between transparent and opaque. This design provides ultrathin optical focusing capabilities by serving as a diffractive lens in which the light transmitted through the transparent zones constructively interferes at the desired focal spot. When fabricated with nonlinear materials, zone plates have demonstrated nonlinear functionality that has been used for applications such as nonlinear imaging and frequency conversion [12–15]. However, their potential for optical limiting has not been previously investigated.

In this work, we propose a nonlinear zone plate that can act as an ultrathin optical limiter. By embedding nonlinear saturable absorbing materials [16] in alternating zones of a dielectric slab, we can achieve a nonlinear zone plate in which the transparency of the embedded regions changes with the intensity of the incident light (Fig. 1). This design allows the zone plate to focus low intensity light while transmitting high intensity as a planar wavefront without a focal point. We computationally design and analyze the device using finite-difference time-domain simulations coupled to Maxwell-Bloch equations that model the saturable absorbing materials. Our simulations of the device demonstrate strong optical limiting at the focal spot of the lens with a threshold intensity as low as 0.45 kW/cm$^2$, which is several orders of magnitude lower than current flat-optics-based optical limiters [4–11]. The zone plate can be optimized for different operating frequencies and device sizes by simply changing the saturable absorbing material and number of zones. With this flexible design capability and low threshold intensity, this nonlinear zone plate approach enables ultrathin optical limiters for miniaturized optical systems.

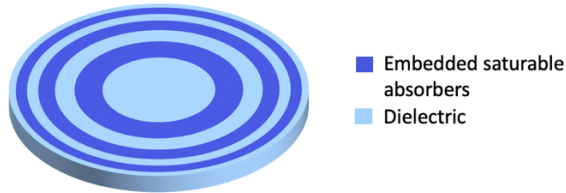

Fig. 1. A schematic of the proposed zone plate optical limiter.

## 2. Proposed Optical Limiting Mechanism

Figure 1 shows a schematic of the proposed nonlinear zone plate. The plate is composed of a transparent dielectric slab that features alternating concentric regions embedded with a saturable absorbing material. At low intensity light, these embedded zones are opaque due to the strong absorption of the saturable absorbers. Therefore, by controlling the widths of these opaque zones, we can achieve constructive interference of the incident light that transmits through the unembedded regions to generate a tight optical focus, enabling the device to act as a diffractive lens (Fig. 2(a)) [17]. However, at high intensity, the saturable absorbers cannot absorb light, which causes the embedded regions to become transparent. In this case, the zone plate behaves as a homogenous dielectric slab that exhibits no focusing capability, functioning as an optical limiter (Fig. 2(b)). Similar concepts have been previously explored in conventional lenses using thermo-optic nonlinearities [1], but these structures were composed of thick bulk optics and featured a high optical limiting threshold due to the weak thermo-optic effect. The proposed zone plate design instead uses a stronger nonlinearity based on saturable absorption to achieve optical limiting functionality at significantly lower power.

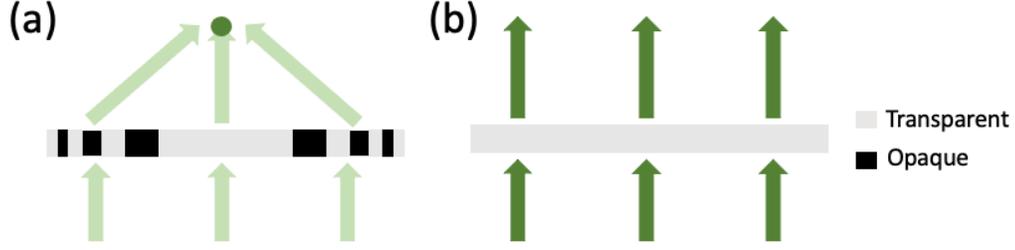

Fig. 2. (a) At low intensity incident light, the regions embedded with saturable absorbers function as opaque zones as they absorb light, while the unembedded regions function as transparent zones. Light transmitted from the transparent zones can constructively interfere at the desired focal length to generate a focal spot. (b) With high intensity light above a certain threshold, the saturable absorbers become saturated and do not absorb light. As a result, the embedded regions function as transparent zones, causing the device to act as a homogeneous thin-film dielectric that transmits light as a planar wave.

## 3. Saturable absorber model

In order to computationally design and numerically simulate the device, we use a numerical model for the saturable absorbing regions. In this model, we treat the saturable absorbing material as an ensemble of homogeneously broadened two-level atoms, with electric susceptibility $\chi_{SA}$, which are embedded in a bulk dielectric substrate, with susceptibility $\chi_D$. The total susceptibility of the embedded region is therefore given by: $\chi_{EM} = \chi_D + \chi_{SA}$, which allows us to simulate the optical response at varying intensities of light. This model is applicable to many saturable absorbing materials, including quantum dots [18], rare-earth ions [19], and low dimensional materials [20]. The response of the saturable absorbers to an incident electric field is described by the Maxwell Bloch equations [21], which allow us to calculate $\chi_{SA}$:

$$\frac{dN_{11}}{dt} = -\frac{i\Omega}{2}(\chi_{SA}^* - \chi_{SA}) + \gamma N_{22} \qquad (1)$$

$$\frac{dN_{22}}{dt} = \frac{i\Omega}{2}(\chi_{SA}^* - \chi_{SA}) - \gamma N_{22} \qquad (2)$$

$$\frac{d\chi_{SA}}{dt} = -(\beta + i\Delta)\chi_{SA} - \frac{i\Omega}{2}(N_{22} - N_{11}) \qquad (3)$$

In the above equations, $N_{11}$ and $N_{22}$ are the ground and excited state population densities of the two-level atoms, respectively, which add up to the total density ($N$) of the embedded two-level atoms in the dielectric slab ($N_{11} + N_{22} = N$). We define $\Omega = \mu E/\hbar$ as the Rabi frequency, where $\mu$ is the transition matrix element of the two-level atom, $E$ is the electric field amplitude, and $\hbar$ is the reduced Planck constant. We define the detuning frequency as $\Delta = \omega_0 - \omega$, where $\omega_0$ is the resonant frequency of the two-level atoms and $\omega$ is the frequency of the incident light. The atomic decay rate is given by $\gamma$, which is defined as $\gamma = \gamma_{non} + \gamma_{rad}$, where $\gamma_{non}$ is the nonradiative decay rate and $\gamma_{rad} = \mu^2\omega^3/3\pi\hbar\varepsilon_0(1 + \chi_D)c^3$ is the radiative decay rate. Finally, $\beta = \gamma/2 + 1/T_2$ where $T_2$ is the dipole dephasing time.

## 4. Device design

*4.1 Numerical simulations of the saturable absorbing material*

Although the device design in Fig. 1 is broadly applicable to many materials, we need to select a specific dielectric and saturable absorber to computationally design the proposed optical limiter. We choose InAs quantum dots as the saturable absorber, which emit at 900 nm, and GaAs as the dielectric ($\chi_D = 11.89$), which sits on a SiO$_2$ substrate, as shown in Fig. 3. To simulate the optical response of the quantum dots, we need to calculate $\chi_{SA}$. At room temperature we can approximate the quantum dots as a homogenously broadened ensemble because their homogenous linewidth is on the same order as their inhomogeneous linewidth [22,23]. This approximation allows us to apply Eq. (1)-(3), which assume a homogenously distributed ensemble, to calculate $\chi_{SA}$.

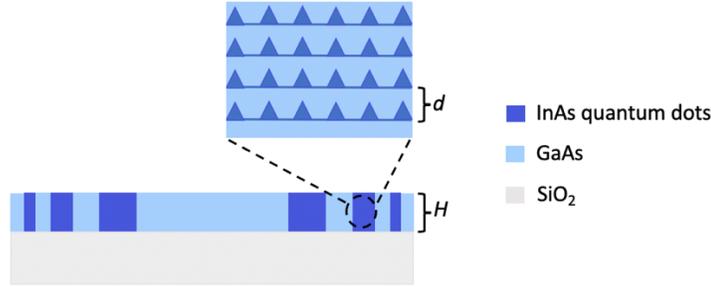

Fig. 3. To computationally design the optical limiter, we specifically consider GaAs as the dielectric material, which sits on a SiO$_2$ layer. For the saturable absorber, we choose InAs quantum dots, which are grown in multiple stacked layers separated by distance *d*.

The value of $\chi_{SA}$ also depends on the density *N* of the embedded InAs quantum dots. Quantum dots are typically grown in layers that are separated by a distance *d*, as illustrated in Fig. 3. Each layer has a two-dimensional quantum dot density of $N_{planar}$, which is typically determined by the material growth parameters. The density of the quantum dots can therefore be defined as $N = N_{planar}/d$. Based on current epitaxial growth techniques, we calculated the quantum dot density using values of $N_{planar} = 7 \times 10^{10}$ cm$^{-2}$ [24,25] and *d* = 14 nm [26], which results in $N = 5 \times 10^{16}$ cm$^{-3}$.

To calculate $\chi_{SA}$ and simulate the optical response of the InAs quantum dots and GaAs, we incorporate the Maxwell-Bloch equations into numerical finite-difference time-domain simulations (Lumerical FDTD Solutions) using the method described by Shih-Hui Chang and Allen Taflove [27]. For all simulations, we set the decay rates of the quantum dots to their room-temperature values of $\gamma_{rad} = 1$ GHz [21,28] and $\gamma_{non} = 1$ GHz [29,30], and the dephasing time to $T_2 = 300\ fs$ [23]. More detailed description of the simulations can be found in the supplementary material.

*4.2 Zone Plate Design*

To achieve a diffractive lens, the embedded and unembedded regions of the zone plate must be opaque and transparent at low intensity light, respectively, which is controlled by the thickness of the device. Therefore, we first simulate the transmission coefficients of these regions at different thicknesses of the GaAs layer ($H$) at low intensity light, where the InAs quantum dots have a strong absorption. The figure of merit to determine the extent of the absorption of the InAs quantum dots is its saturation intensity: $I_{sat} = \hbar\omega^3\gamma(\beta^2 + \Delta^2)/6\sqrt{(1+\chi_D)}\pi c^2\beta\gamma_{rad} = 0.095 kW/cm^2$. Therefore, we use $I = I_{sat} \times 10^{-8}$ for the low intensity light, where the quantum dots are well below saturation. In the simulation, we set the wavelength of the incident plane wave to 900 nm, which is resonant with the quantum dots and therefore should result in the high absorption necessary to create opaque zones.

Figure 4 shows the calculated transmission coefficients of the bare GaAs regions and the areas embedded with InAs quantum dots as a function of $H$. The simulated transmission coefficient of the embedded region decreases from 0.96 to ~0 when $H$ increases from 0 to 0.8 $\mu m$. For the unembedded regions, the transmission coefficient oscillates between 0.96 and 0.35 with a periodicity of 125 nm (corresponding to half of the light's wavelength inside the GaAs), which can be explained by Fabry-Pérot resonance [31,32]. Based on these results, we set $H =$ 0.5 $\mu m$ for the initial design of the zone plate since at this thickness the embedded and unembedded regions of the device can be considered opaque and transparent at low intensity light, as evidenced by the corresponding transmission coefficients of 0.004 and 0.9 (black dashed line in Fig. 4).

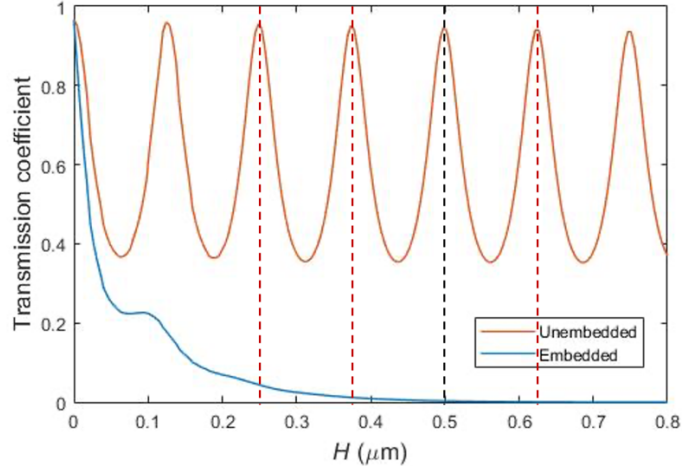

Fig. 4. The transmission coefficients of the embedded and unembedded regions of the zone plate as a function of the thickness ($H$) of the GaAs layer. The transmission coefficient of the embedded regions decreases with $H$ due to the increasing absorption of the saturable absorbers, while for the unembedded regions the transmission oscillates as a result of Fabry-Pérot resonance.

We then arrange the opaque and transparent zones of the device to achieve constructive interference of the transmitted light at the focal point in the low intensity regime. As shown in Fig. 5, the zone plate consists of concentric regions of varying radii that alternate between transparent and opaque. Constructive interference at a focal length $f$ is realized when the beams of light passing through the transparent zones reach the selected focal point with phases that differ by less than $\pi$. This can be achieved if the optical path length from the focal spot to the $n^{\text{th}}$ concentric circle is $f + n\lambda/2$ (Fig. 5) [33]. As a result, we can calculate the desired radius $r_n$ of the $n^{\text{th}}$ ring by:

$$r_n = \sqrt{n\lambda f + \frac{n^2\lambda^2}{4}}, n \leq n_0 \qquad (4)$$

where $n_0$ is the total number of zones in the zone plate. The values of $f$ and $n_0$ in Eq. (4) can be chosen to obtain different focal lengths and sizes of the zone plate depending on the application. Here we choose $f = 100$ $\mu m$ and $n_0 = 30$, and calculate the exact values of each $r_n$ using Eq. (4), which can be found in Table S1 in the supplementary material. Thus under these conditions, with the InAs quantum dots embedded in the even numbered zones, we can create a nonlinear zone plate.

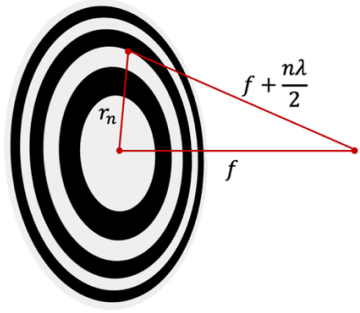

Fig. 5. In a zone plate, the optical path length from the $n^{\text{th}}$ ring to the desired focal spot is described by $f + n\lambda/2$. By setting the even zones to be opaque by embedding InAs quantum dots and the odd zones to be transparent (i.e., bare dielectric), the zone plate can achieve constructive interference at the focal spot.

## 5. Optical limiter analysis

To investigate the optical limiting effect of this nonlinear zone plate ($H = 0.5$ $\mu m$ and $n_0 = 30$), we first simulate its lensing behavior at an intensity far below the saturation of the InAs quantum dots ($I = I_{sat} \times 10^{-8}$). We define a coordinate system such that the cross-section of the zone plate lies along the $x$ axis with the center of its surface at the origin, as shown in Fig. 6(a). We excite the zone plate with a continuous plane wave at normal incidence (i.e., along the $y$ axis) and set the wavelength of the incident beam to 900 nm (on-resonance with the InAs quantum dots). Figure 6(b) shows the simulated far field intensity above the zone plate, in which the lens achieves a tight focal spot at a position of (x, y) = (0, 100) $\mu m$.

We then simulate the zone plate's lens behavior when the incident intensity is well above the saturation of the quantum dots ($I = I_{sat} \times 10^2$) while keeping the other simulation parameters the same. Under these conditions, we expect the embedded zones to be transparent. Figure 6(c) plots the simulated far field intensity above the device. The intensity no longer exhibits a tight focal point but instead is uniformly distributed over the far field region (the oscillatory patterns are due to diffraction from the edges of the device). This result shows that for high intensity light the zone plate transforms from a flat lens to a thin, uniform dielectric slab that is unable to focus light.

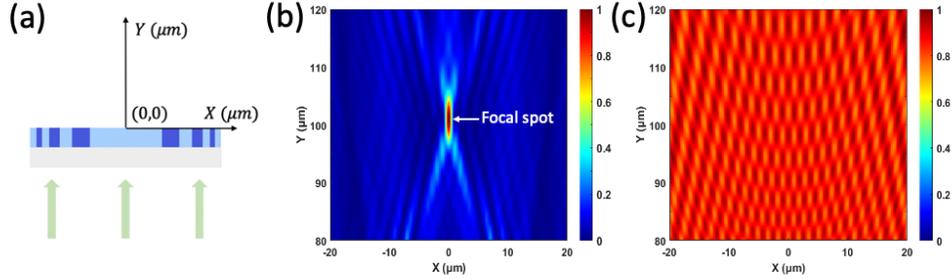

Fig. 6. (a) We simulate the far field intensity of the light that transmits through the zone plate under both low ( $I = I_{sat} \times 10^{-8}$ ) and high ( $I = I_{sat} \times 10^{2}$ ) intensities, with the zone plate sitting at y = 0. (b) At low intensity, light is focused at (x, y) = (0, 100) μm. (c) At high intensity, instead of a focal spot, we observe only patterns that result from the diffraction of light at the edges of the device, indicating the light is transmitted in a plane wave instead of focused.

To determine the optical limiting threshold of the device, we plot the output intensity at the focal spot position indicated in Fig. 6 (b) as a function of the input intensity. Figure 7 (a) shows the results of the calculation. The output intensity increases as we increase the input intensity from 0 to 0.45 kW/cm², then slightly decreases until 1 kW/cm², and finally barely increases even as the input approaches 1.8 kW/cm². These results show this nonlinear zone plate ($H$ = 0.5 μm and $n_0 = 30$) has a strong optical limiting effect and can restrict the output intensity when the input exceeds a threshold of 0.45 kW/cm².

We also calculate the transmission coefficient and focusing efficiency of the zone plate as a function of the input intensity. The transmission coefficient is defined as the fraction of incident light that transmits through the device. Meanwhile, the focusing efficiency describes the ability of the zone plate to focus light, which is defined by the fraction of the incident light that passes through a linear aperture in the focal plane that has a length of three times the full-width at half maximum of the focal spot [34-35]. As shown in Fig. 7(b), the transmission efficiency increases with increasing power due to the reduced absorption of the InAs quantum dots as they saturate. Despite the increase in the transmission efficiency, the focusing efficiency of the zone plate decreases from 11.5% to 4% as the input intensity increases from 0 to 1.8 kW/cm² due to the transformation of the device from a lens to a thin dielectric slab. This drop in focusing efficiency provides the nonlinear zone plate with a strong optical limiting effect.

We note that $H$ = 0.5 μm is just a sample thickness that can be used for the zone plate design. Optical limiting can be achieved at different device thicknesses as long as there is a high contrast between the transmission coefficients of the embedded and unembedded regions at low intensity light. For example, besides the thickness used in the initial design ($H$ = 0.5 μm), we investigated three additional values (0.25, 0.375, and 0.625 μm). At these values of $H$, the unembedded regions have the same maximum transmission coefficient at low intensity, while the embedded regions have decreasing values of 0.04, 0.01 and 0.001, respectively, as shown by the red dashed lines in Fig. 4. To investigate how these values of $H$ impact the optical limiting behavior, we simulate the output intensity as a function of the input intensity while keeping the other parameters fixed (Fig. 7(c)). The threshold intensity is indicated by the dashed lines for each device, which increases with $H$ as a higher input intensity is required to saturate a thicker region of quantum dots.

Figure 7(c) also shows that zone plates with a higher value of $H$ have a steeper slope in their output curve when the input intensity is less than the limiting threshold. A steeper curve corresponds to a higher focusing efficiency, indicating that more light is focused by thicker devices. This trend can be explained by the lower transmission coefficient of the embedded regions at higher thickness (i.e., they are opaquer). Additionally, after the threshold intensity, higher values of $H$ display flatter or even decreasing slopes in the output curves, which indicates a stronger optical limiting effect. However, as we further increase the device thickness, improvements in the focusing efficiency and optical limiting effect will gradually saturate when the transmission coefficient of the embedded region is so close to 0 as to be considered fully opaque. For example, as shown in Fig. 7 (c), we see little improvement in the optical limiting behavior when we further increase $H$ from 0.5 $\mu m$ (yellow) to 0.625 $\mu m$ (purple), which correspond to very low transmission coefficients of 0.004 and 0.001, respectively, in the embedded regions of the device.

The number of zones $n_0$ in the nonlinear zone plate can also affect the optical limiting performance, as shown in Fig. 7(d). With $H$ fixed at $H = 0.5$ $\mu$m, zone plates with a higher $n_0$ have higher output intensity for a given input below the threshold, because more power can be transmitted and focused by larger devices. A larger $n_0$ also contributes to a better optical limiting behavior as evidenced by a flatter or decreasing curve above the threshold. However, the threshold intensity does not vary at different $n_0$ and only depends on $H$.

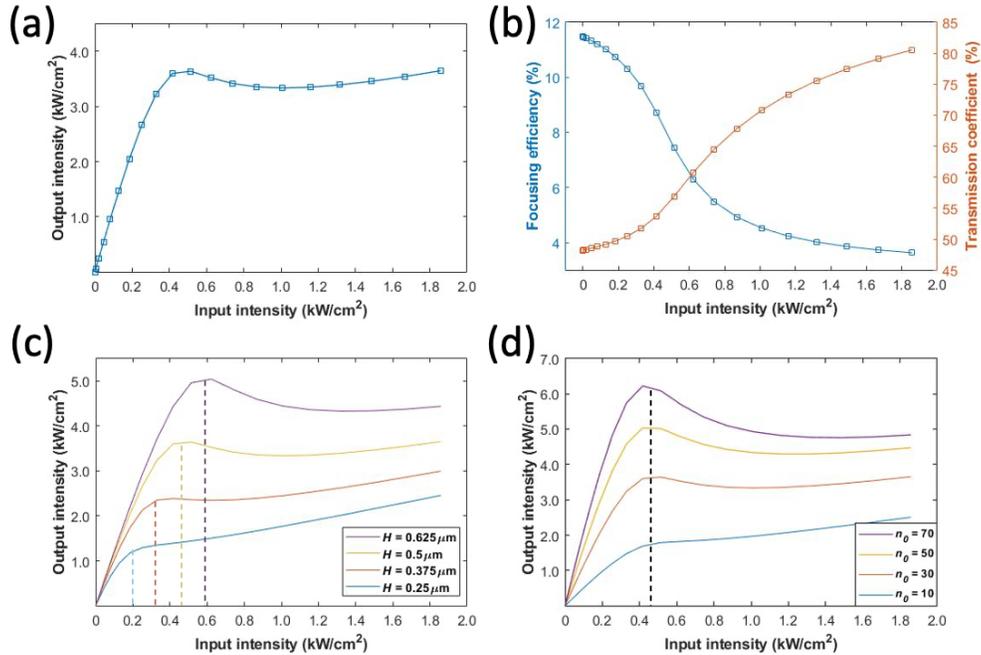

Fig. 7. (a) The output intensity of the zone plate ($H = 0.5$ $\mu$m and $n_0 = 30$) plateaus at a threshold intensity of 0.45 kW/cm$^2$, indicating a strong optical limiting effect. (b) The transmission coefficient increases with input intensity due to the decreasing absorption of the InAs quantum dots. Meanwhile, the focusing efficiency decreases with input intensity due to the device's transition from a lens to a thin dielectric slab, which provides a strong optical limiting effect. (c) The output intensity as a function

of the input intensity for zone plates with the same $n_0$ (30) but different values of $H$ (0.25, 0.375, 0.50, and 0.625 $\mu m$). A thicker device has a higher threshold intensity. Additionally, at larger $H$, a zone plate has a higher focusing efficiency and better limiting effect. (d) The output intensity as a function of the input intensity for zone plates with the same $H$ (0.5 $\mu m$) but different $n_0$ (10, 30, 50, and 70). The threshold intensity does not vary at different $n_0$ and only depends on $H$. With a larger $n_0$, the zone plate has a better limiting effect (i.e., the output curve above the threshold is flatter or decreasing in intensity).

## 6. Conclusions

In this work, we show a nonlinear zone plate composed of a dielectric embedded with alternating zones of a saturable absorbing material can act as a low-power optical limiter. Using InAs quantum dots as an example of the saturable absorber, we computationally designed an ultrathin optical limiter with a threshold intensity as low as 0.45 kW/cm². This threshold is several orders of magnitude smaller than those observed in current flat optics-based optical limiters [4-11]. Our nonlinear zone plate design could enable ultra-thin optical limiters and potentially find broad application where nonlinear focusing ability is needed, such as optical signal processing [36-38] and optical machine learning [39].

### Funding

The authors would like to acknowledge funding from an ONR grant (grant #N000142012551).

### Disclosures

The authors declare no conflicts of interest.

### Supplemental document

See Supplement 1 for supporting content.

### References

1. H. Nadjari, F. Hajiesmaeilbaigi, and A. Motamedi, "Thermo optical response and optical limiting in Ag and Au nanocolloid prepared by laser ablation," Laser Phys. **20**(4), 859–864 (2010).
2. B. L. Justus, Z. H. Kafafi, and A. L. Huston, "Excited-state absorption-enhanced thermal optical limiting in C_60," Opt. Lett. **18**(19), 1603 (1993).
3. B. L. Justus, A. L. Huston, and A. J. Campillo, "Broadband thermal optical limiter," Appl. Phys. Lett. **63**(11), 1483–1485 (1993).
4. H. Qian, S. Li, Y. Li, C. F. Chen, W. Chen, S. E. Bopp, Y. U. Lee, W. Xiong, and Z. Liu, "Nanoscale optical pulse limiter enabled by refractory metallic quantum wells," Sci. Adv. **6**(20), 1–7 (2020).
5. S. A. Mann, N. Nookala, S. Johnson, A. Mekkawy, J. F. Klem, I. Brener, M. Raschke, A. Alù, and M. A. Belkin, "Ultrafast optical switching and power limiting in intersubband polaritonic metasurfaces," Opt. InfoBase Conf. Pap. **Part F182**-(5), (2020).
6. S. Guddala and S. A. Ramakrishna, "Optical limiting by nonlinear tuning of resonance in metamaterial absorbers," Opt. Lett. **41**(22), 5150 (2016).
7. X. Zhao, J. Zhang, K. Fan, G. Duan, G. D. Metcalfe, M. Wraback, X. Zhang, and R. D. Averitt, "Nonlinear terahertz metamaterial perfect absorbers using GaAs [Invited]," Photonics Res. **4**(3), A16 (2016).
8. B. Slovick, L. Zipp, and S. Krishnamurthy, "Indium phosphide metasurface with enhanced nonlinear absorption," Sci. Rep. **7**(1), 1–7 (2017).
9. S. Jafar-Zanjani, J. Cheng, V. Liberman, J. B. Chou, and H. Mosallaei, "Large enhancement of third-order nonlinear effects with a resonant all-dielectric metasurface," AIP Adv. **6**(11), (2016).
10. A. Howes, Z. Zhu, D. Curie, J. R. Avila, V. D. Wheeler, R. F. Haglund, and J. G. Valentine, "Optical Limiting Based on Huygens' Metasurfaces," Nano Lett. **20**(6), 4638–4644 (2020).
11. L. Hsu and A. Ndao, "Diffraction-limited broadband optical," Opt. Lett. **46**(6), 1293–1296 (2021).
12. E. Minerbi, S. Keren-Zur, and T. Ellenbogen, "Nonlinear Metasurface Fresnel Zone Plates for Terahertz Generation and Manipulation," Nano Lett. **19**(9), 6072–6077 (2019).
13. N. Charipar, P. Johns, R. J. Suess, H. Kim, J. Geldmeier, S. Trammell, K. Charipar, J. Naciri, A. Piqué, and J. Fontana, "Light tunable plasmonic metasurfaces," Opt. Express **28**(15), 22891 (2020).
14. O. Manela and M. Segev, "Nonlinear diffractive optical elements," Opt. InfoBase Conf. Pap. **15**(17), 10863–

 10868 (2007).
15. N. Segal, S. Keren-Zur, N. Hendler, and T. Ellenbogen, "Controlling light with metamaterial-based nonlinear photonic crystals," Nat. Photonics **9**(3), 180–184 (2015).
16. M. Hercher, "An Analysis of Saturable Absorbers," Appl. Opt. **6**(5), 947 (1967).
17. F. Capasso, "The future and promise of flat optics: A personal perspective," Nanophotonics **7**(6), 953–957 (2018).
18. X. Wang, Y. J. Zhu, C. Jiang, Y. X. Guo, X. T. Ge, H. M. Chen, J. Q. Ning, C. C. Zheng, Y. Peng, X. H. Li, and Z. Y. Zhang, "InAs/GaAs quantum dot semiconductor saturable absorber for controllable dual-wavelength passively Q-switched fiber laser," Opt. Express **27**(15), 20649 (2019).
19. S. Stepanov, "Dynamic population gratings in rare-earth-doped optical fibres," J. Phys. D. Appl. Phys. **41**(22), (2008).
20. Z. Li, C. Pang, R. Li, and F. Chen, "Low-dimensional materials as saturable absorbers for pulsed waveguide lasers," J. Phys. Photonics **2**(3), 031001 (2020).
21. D. Sridharan and E. Waks, "All-optical switch using quantum-dot saturable absorbers in a DBR microcavity," IEEE J. Quantum Electron. **47**(1), 31–39 (2011).
22. H. Taleb and K. Abedi, "Homogeneous and inhomogeneous broadening effects on static and dynamic responses of quantum-dot semiconductor optical amplifiers," Front. Optoelectron. **5**(4), 445–456 (2012).
23. P. Borri, W. Langbein, J. Mørk, J. M. Hvam, F. Heinrichsdorff, M. H. Mao, and D. Bimberg, "Dephasing in InAs/GaAs quantum dots," Phys. Rev. B - Condens. Matter Mater. Phys. **60**(11), 7784–7787 (1999).
24. K. Akahane, N. Yamamoto, and M. Tsuchiya, "Highly stacked quantum-dot laser fabricated using a strain compensation technique," Appl. Phys. Lett. **93**(4), 3–6 (2008).
25. D. Guimard, M. Nishioka, S. Tsukamoto, and Y. Arakawa, "High density InAs/GaAs quantum dots with enhanced photoluminescence intensity using antimony surfactant-mediated metal organic chemical vapor deposition," Appl. Phys. Lett. **89**(18), 1–4 (2006).
26. T. Srinivasan, P. Mishra, S. K. Jangir, R. Raman, D. V. Sridhara Rao, D. S. Rawal, and R. Muralidharan, "Molecular Beam Epitaxy growth and characterization of silicon - Doped InAs dot in a well quantum dot infrared photo detector (DWELL-QDIP)," Infrared Phys. Technol. **70**, 6–11 (2015).
27. S.-H. Chang and A. Taflove, "Finite-difference time-domain model of lasing action in a four-level two-electron atomic system," Opt. Express **12**(16), 3827 (2004).
28. H. Fujita, K. Yamamoto, J. Ohta, Y. Eguchi, and K. Yamaguchi, "In-plane quantum-dot superlattices of InAs on GaAsSb/GaAs(001) for intermediate band solar-cells," Conf. Rec. IEEE Photovolt. Spec. Conf. (001), 002612–002614 (2011).
29. C. H. Lin, H. S. Lin, C. C. Huang, S. K. Su, S. D. Lin, K. W. Sun, C. P. Lee, Y. K. Liu, M. D. Yang, and J. L. Shen, "Temperature dependence of time-resolved photoluminescence spectroscopy in InAs/GaAs quantum ring," Appl. Phys. Lett. **94**(18), 13–16 (2009).
30. W. Yang, R. R. Lowe-Webb, H. Lee, and P. C. Sercel, "Effect of carrier emission and retrapping on luminescence time decays in InAs/GaAs quantum dots," Phys. Rev. B - Condens. Matter Mater. Phys. **56**(20), 13314–13320 (1997).
31. P. Ragulis, A. Matulis, and Ž. Kancleris, "Fabry-Perot resonances in a symmetric three layer structure," J. Appl. Phys. **118**(12), (2015).
32. N. Calander, "Surface plasmon-coupled emission and fabry - Perot resonance in the sample layer: A theoretical approach," J. Phys. Chem. B **109**(29), 13957–13963 (2005).
33. J. E. Garrett and J. C. Wiltse, "Fresnel zone plate antennas at millimeter wavelengths," Int. J. Infrared Millimeter Waves **12**(3), 195–220 (1991).
34. Z. P. Zhuang, R. Chen, Z. Bin Fan, X. N. Pang, and J. W. Dong, "High focusing efficiency in subdiffraction focusing metalens," Nanophotonics **8**(7), 1279–1289 (2019).
35. A. Arbabi, Y. Horie, A. J. Ball, M. Bagheri, and A. Faraon, "Subwavelength-thick lenses with high numerical apertures and large efficiency based on high-contrast transmitarrays," Nat. Commun. **6**(May), 2–7 (2015).
36. H. Liang, Q. Lin, X. Xie, Q. Sun, Y. Wang, L. Zhou, L. Liu, X. Yu, J. Zhou, T. F. Krauss, and J. Li, "Ultrahigh Numerical Aperture Metalens at Visible Wavelengths," Nano Lett. **18**(7), 4460–4466 (2018).
37. K. Kodate, E. Tokunaga, Y. Tatuno, J. Chen, and T. Kamiya, "Efficient zone plate array accessor for optoelectronic integrated circuits: design and fabrication," Appl. Opt. **29**(34), 5115 (1990).
38. M. M. Moeini and D. L. Sounas, "Discrete space optical signal processing," Optica **7**(10), 1325 (2020).
39. X. Lin, Y. Rivenson, N. T. Yardimci, M. Veli, Y. Luo, M. Jarrahi, and A. Ozcan, "All-optical machine learning using diffractive deep neural networks," Science (80-. ). **361**(6406), 1004–1008 (2018).

# A Low Power Threshold, Ultrathin Optical Limiter Based on a Nonlinear Zone Plate: supplemental document

**Finite-difference time-domain electromagnetic simulations in the presence of two-level atomic systems:**

When optically pumped for sufficiently long, the quantum dots reach a steady state in which the electron population densities ($N_{11}$ and $N_{22}$) and the electric susceptibility $\chi_{SA}$ in Eq. (1)-(3) reach a constant steady-state value. The required pumping time to reach this steady state should be on the order of a few excited lifetimes. Based on the atomic decay rate $\gamma$, the excited state lifetime of the InAs quantum dots is given by $1/\gamma = 0.5$ ns. However, nanoseconds of nonlinear simulations can be very time-consuming and impractical to run. To overcome this limit, we note that we are only interested in simulating the steady-state condition of the quantum dots rather than the initial dynamic optical response. Therefore, we can solve Eq. (1)-(3) in the steady state (when the derivatives equal zero) to obtain the following expressions for $\chi_{SA}$ and $N_{22}$:

$$\chi_{SA} = \frac{i3\pi(1+\chi_D)c^3\gamma N}{2\omega^3(\beta+i\Delta)} \frac{1}{1+\frac{I}{I_{sat}}} \tag{S1}$$

$$N_{22} = \frac{N\, I/I_{sat}}{2(1+I/I_{sat})} \tag{S2}$$

where $I$ is the intensity of the incident light and the saturation intensity $I_{sat}$ is given by:

$$I_{sat} = \frac{\hbar\omega^3\gamma(\beta^2+\Delta^2)}{6\sqrt{(1+\chi_D)}\pi c^2\beta\gamma_{rad}} = 0.095\,kW/cm^2 \tag{S3}$$

Eq. (S1) indicates that as long as the value of $\gamma/2 \ll 1/T_2$ such that $\beta \approx 1/T_2$, $\chi_{SA}$ only depends on the product $\gamma N$. We can therefore define two new variables, $\bar{\gamma} = \gamma\alpha$ and $\bar{N} = N/\alpha$, for any value of $\alpha$. In the steady state, the simulation results using $\bar{\gamma}$ and $\bar{N}$ would yield the same results as for $\gamma$ and $N$. We can therefore reduce the simulation time by setting $\alpha$ to a large value, provided that we later convert back to the original values of the atomic decay rate and density. However, we note that $\alpha$ cannot be arbitrarily large because at some point the original approximation $\beta \approx 1/T_2$ will no longer hold. Therefore, in our simulations we set $\alpha$ to 100. This allows us to decrease the simulation time 100-fold by decreasing $1/\gamma$ while maintaining the same steady-state condition.

**Zone plate design:**

In order to achieve a constructive interference at the desired focal length ($f = 100\,\mu$m), we calculate the required values of the radius ($r_n$) of each concentric circle in the zone plate from Eq. (4), as shown in Table S1. below. The zone plate is designed to have 30 zones in total ($n_0 = 30$).

| n | 1 | 2 | 3 | 4 | 5 | 6 | 7 | 8 | 9 | 10 |
|---|---|---|---|---|---|---|---|---|---|---|
| $r_n$ (μm) | 9.50 | 13.45 | 16.49 | 19.06 | 21.33 | 23.39 | 25.30 | 27.07 | 28.75 | 30.34 |
| n | 11 | 12 | 13 | 14 | 15 | 16 | 17 | 18 | 19 | 20 |
| $r_n$ (μm) | 31.85 | 33.30 | 34.70 | 36.05 | 37.36 | 38.62 | 39.86 | 41.06 | 42.23 | 43.37 |
| n | 21 | 22 | 23 | 24 | 25 | 26 | 27 | 28 | 29 | 30 |
| $r_n$ (μm) | 44.49 | 45.59 | 46.66 | 47.71 | 48.75 | 49.77 | 50.77 | 51.76 | 52.73 | 53.69 |

Table S1. The values of $r_n$ of the designed zone plate ($f = 100$ μm, $n_0 = 30$)